\RequirePackage{lineno}
\documentclass[twocolumn]{revtex4}
\usepackage{amsmath}
\usepackage{amssymb}
\usepackage{graphicx}
\usepackage{dcolumn}
\usepackage{bm}

\begin{document}

\title{ Tuning the magnetic configuration of bilayer graphene quantum dot by twisting  }
\author{Ma Luo\footnote{Corresponding author:swym231@163.com}}
\affiliation{School of Optoelectronic Engineering, Guangdong Polytechnic Normal University, Guangzhou 510665, China}

\begin{abstract}

Twistronic has recently attracted tremendous attention because the twisting can engineer the bilayer graphene-like materials into varying types of strongly correlated phases. In this paper, we study the twisting of bilayer graphene (BLG) quantum dots (QDs) with hexagonal shape and zigzag edges. In the untwisted BLG-QDs, the zigzag edges of graphene host spontaneous magnetism with varying magnetic configurations. As a BLG-QD being adiabatically twisted, the quantum state evolves as a function of the twisting angle. If the twisting angle changes across certain critical value, the magnetic configuration of the quantum state sharply changes. For the twisting process with increasing or decreasing twisting angle, the number and value of the critical twisting angles are different. Thus, the twisting process with the twisting angle increasing and decreasing back and forth could enter a hysteresis loop. The twisting of BLG QDs with adatom is also investigated. The tuning features of the magnetic configuration of the twisted BLG-QDs could be applied for graphene-based quantum memory devices.

\end{abstract}

\maketitle

\section{Introduction}

Graphene has been proposed to be exceptional candidate as building block of new-generation electronic integrated systems, because high efficient electronic and heat conductivity can be obtained by the simple hexagonal structure of carbon \cite{Zutic04,WHan14,YuguiYao2011,Motohiko12,YRen16}. The behavior of electron at the Fermi level can be modeled by Dirac Fermion model, which has similar dispersion as light in vacuum, except that the speed is 300 times smaller than light speed \cite{CastroNeto09}. In nanoribbons with finite width, the size effect alters the dispersion of electron near to the Fermi level. The band gap as well as the topological property of armchair nanoribbon is dependent on the width of the nanoribbon \cite{StevenLouie18,StevenLouie21}. On the other hand, spontaneous magnetism of zigzag edge modifies the band structure of zigzag nanoribbon \cite{Mitsutaka96,Hikihara03,Yamashiro03,YoungWooSon06,YoungWoo06,Pisani07,Wunsch08,FernandezRossier08,Jung09,Rhim09,Lakshmi09,Jung09a,Yazyev10,Hancock10,Manuel10,Feldner11,DavidLuitz11,JeilJung11,Culchac11,Schmidt12,Karimi12,Schmidt13,Golor13,Bhowmick13,FengHuang13,Ilyasov13,Carvalho14,Lado14,MichaelGolor14,PrasadGoli16,Baldwin16,Ortiz16,Hagymasi16,Ozdemir16,Friedman17,ZhengShi17,Krompiewski17a,Krompiewski19a,maluo2020}. The band gap is dependent on the magnetic configuration as well as the width of the nanoribbon \cite{Jung10,Soriano12,XiaoLong18}. In graphene quantum dot (QD) with at least three zigzag edges, the spontaneous magnetized states with varying magnetic configurations can be formed with high stability \cite{Rossier07,WeiHu19}. In bilayer graphene (BLG) zigzag nanoribbon, the magnetic configuration can be tuned by gated voltage in perpendicular direction, which in turn tune the band structure \cite{maluo21}. Combination of gated voltage in perpendicular direction and electric field in horizontal direction can flip the nanoribbon among different magnetic configurations \cite{maluo21a,Tixuan22}.

The BLG can be tuned by twisting the relative angle between the two graphene layers \cite{Santos07,Morell10,Trambly10,Bistritzer11}. The matching and coupling between the Dirac Fermion on the two graphene layers form van-Hove singularity on the band structure, so that ultra-flat band of electron can be engineered \cite{Morell10}. As a result, the dynamical hopping term become much smaller than the interaction terms, which turns the twisted BLGs into strongly correlated materials. Novel superconductivity and ferroelectric phase are found in these fascinating materials \cite{Cao18,Bernevig21,Ahn18}. The twisting can also tune the quantum state of finite size BLG QDs \cite{Mirzakhani20,Naik20,XiaoZhou21,Bucko21,Rakhmanov22,YunhuaWang22}, which in turn modifies the electric dipole polarizability, optical excitation selection rule, and the Landau level.

In this paper, we studied the spontaneous magnetism in twisted BLG QDs with hexagonal shape and six zigzag edges in each layer. Mean field method is applied to self-consistently solve the tight binding model with Hubbard interaction. There are twelve zigzag edges in total, so that $2^{12}$ different magnetic configurations could appear in the untwisted BLG QDs. Because of the symmetry of the structure, some configurations are equivalent. In additional, some configurations are not stable. At the ground state, the magnetic moments of the terminations at each zigzag edges is anti-parallel to that at the intra-layer neighboring zigzag edges, and parallel to that at the inter-layer neighboring zigzag edge. The quasi-stable excited states can be obtained by flipping the magnetic moments at one or more than one of the zigzag edges. As the untwisted BLG QD initially being in the ground state, when the twisting angle is increased from zero to a finite value, the distribution of the magnetic moments smoothly changes as the twisting angle increases, and sharply changes as the twisting angle reaches certain critical values. The value of the critical twisting angle are numerically calculated. Similar phenomenon occurs to the reversed twisting process, whose twisting angle decrease from a finite value to zero, except that the value of the critical twisting angle are different. Thus, if the BLG QDs are twisted back and forth, the quantum state could enter a hysteresis loop.

Additional presence of adatom in graphene can also induce magnetism \cite{Nokelainen19,Jeongsu19,Wellnhofer19,SungWoo19,Noori20,YuZhang20}. The spin splitting of the localized quantum state near to the adatom induce spin dependent scattering of carrier, which can be applied in spintronic devices. Other form of carbon materials with adatom, such as hydrogenated buckyball fullerene ($C_{60}$), also host spontaneous magnetism with large magnetic moment \cite{XiaoBaoYang21}. In this paper, the twisting properties of the BLG QDs with one adatom in the middle is investigated. In the hysteresis loop, the magnetic moments near to the adatom oscillate between positive and negative direction. The bistability behavior of the magnetic moments could be applied as nano-scale quantum memoriser.

The paper is organized as the following: In Sec. II, the theoretical model of the twisted BLG QDs is described. In Sec. III, the numerical result of the twisted BLG QDs with and without adatom is discussed. In Sec IV, the conclusion is given.

\section{Theoretical model}

\begin{figure*}[tbp]
\scalebox{0.73}{\includegraphics{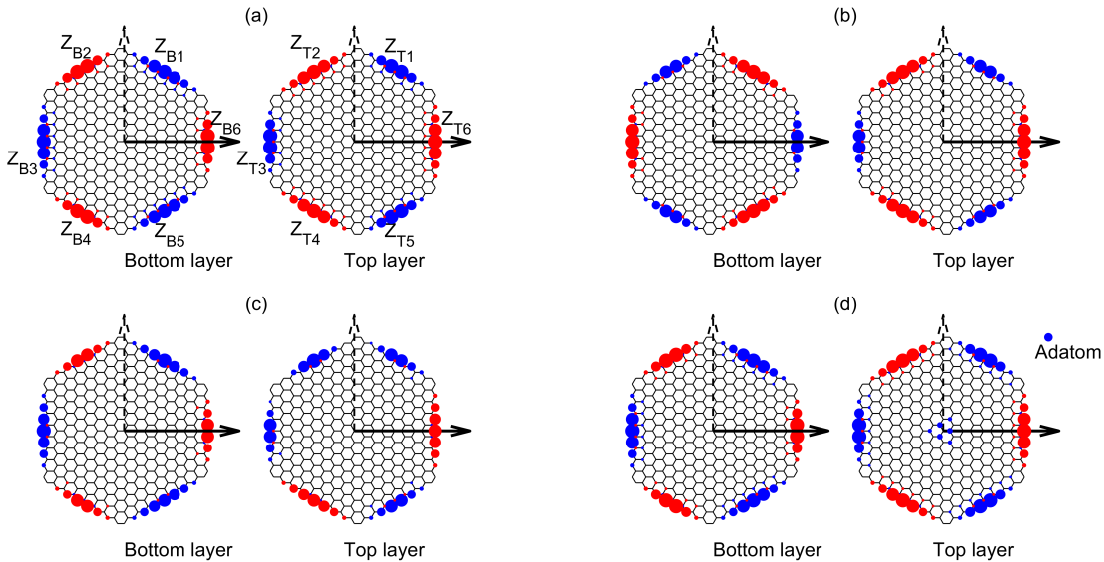}}
\caption{ Lattice structure and spatial distribution of magnetic moments of the untwisted BLG QDs without adatom in (a-c), and with adatim in (d). For better visualization, in each sub-figure, the lattice structure of the top and bottom layers are shifted to the right and left, respectively. The lattice site of the adatom in (d) is shifted to the top-right corner of the sub-figure. The distributions of magnetic moments of the ground state in (a) and (d), the first quasi-stable excited state in (b) and the AFM-AFM$^{-}$-0 state in (c) are indicated by the size and color of the marker in each lattice site, with the size representing the magnitude, the color representing the direction. The blue and red markers represent upward and downward magnetic moment, respectively. }
\label{figure_1}
\end{figure*}

The structure of the BLG QD is indicated in Fig. \ref{figure_1}(a). For better visualization, the drawing of the top and bottom layers' lattice structure are horizontally shifted to the right and left half of the sub-figure, respectively. Because the BLG QDs consisted of AB stack BLGs have $D_{3d}$ symmetry, each graphene layer is not regular hexagon, but hexagon with alternating length of zigzag edge. The zigzag edges at top and bottom layers are designated in counter-clockwise sequence as $Z_{Ti}$ and $Z_{Bi}$, respectively, with $i\in[1,6]$, which are marked in Fig. \ref{figure_1}(a). The zigzag edges $Z_{T(2j-1)}$ and $Z_{B(2j)}$ with $j\in[1,3]$ have $2N$ zigzag terminations, and the other zigzag edges have $2N+1$ zigzag terminations, with $N$ being integer that represent the size of the BLG QDs. The x-y coordinate origin locates at one of the AB stacking sites in the center of the BLG QDs. The solid and dashed arrows indicate the x and y axis of the lattice structure, respectively. If the BLG is AA stack with regular hexagon in each layer, the BLG QQ have $D_{3h}$ symmetry, and the x-y coordinate origin locates at the hollow center of each graphene layer. Because the AB stack untwisted BLG is more stable than the AA stack counterpart, this paper focus on the twisting of the AB stack BLG QDs.

The tight binding model is applied to simulate the quantum state of the BLG QDs. In each lattice site, the $\pi$ orbit is included in the model. For intra-layer motion, only hopping terms between the nearest neighboring lattice sites are included. In untwisted BLG QDs, the inter-layer motion can be well approximated by including only the hopping terms between the AB stacking pairs of lattice sites. As the BLG QDs being twisted around the center, the lattice site in the center of the top layer remain being on top of the lattice site in the center of the bottom layer; none of the other lattice site in the top layer is exactly on top of any lattice site in the bottom layer. Exceptions occur as the twisting angle equating to one of the magical angles, which bring commensuration to the lattice structures of the two twisted graphene layers. However, only a small portion of lattice sites in the top layer are exactly on top of the lattice sites in the bottom layer. As a result, for inter-layer motion, hopping terms between lattice sites with large distance should be included. Because the strength of the hopping terms is exponentially decaying as function of distance, only the hopping terms between two inter-layer lattice sites with distance smaller than $4a_{C}$ are included, with $a_{C}$ being the length of carbon-carbon bound. The Hamiltonian of the tight binding model can be written as
\begin{equation}
H=\sum_{i,j,s}t(\mathbf{R}_{i},\mathbf{R}_{j})c^{\dag}_{i,s}c_{j,s}+U\sum_{i}\hat{n}_{i,+}\hat{n}_{i,-}
\label{hamiltonian}
\end{equation}
, where $i$ and $j$ are indices of lattice sites, $s=\pm$ represent spin up and down, $t(\mathbf{R}_{i},\mathbf{R}_{j})$ is the hopping term between the lattice sites at $\mathbf{R}_{i}$ and $\mathbf{R}_{j}$, and $c^{\dag}_{i,s}$($c_{i,s}$) is the creation (annihilation) operator of the orbit at the i$^{th}$ lattice site of spin $s$. The hopping terms in the first summation are approximated as \cite{twistTBcoeff1,twistTBcoeff2,twistTBcoeff3,twistTBcoeff4,twistTBcoeff5,twistTBcoeff6}
\begin{equation}
t(\mathbf{R}_{i},\mathbf{R}_{j})=V_{pp\pi}[1-(\frac{\mathbf{d}\cdot\mathbf{e}_{z}}{d})^{2}]+V_{pp\sigma}(\frac{\mathbf{d}\cdot\mathbf{e}_{z}}{d})^{2}
\end{equation}
where
\begin{equation}
V_{pp\pi}=V_{pp\pi}^{0}e^{-\frac{d-a_{C}}{\delta}}
\end{equation}
\begin{equation}
V_{pp\sigma}=V_{pp\sigma}^{0}e^{-\frac{d-d_{0}}{\delta}}
\end{equation}
with $\mathbf{d}=\mathbf{R}_{i}-\mathbf{R}_{j}$ and $d=|\mathbf{d}|$, $\mathbf{e}_{z}=\hat{z}$, $d_{0}=0.335$ nm being the interlayer distance. The other parameters can be obtained by overlap integral of the $\pi$ orbit, which are approximated as $V_{pp\pi}^{0}\approx-2.7$ eV, $V_{pp\sigma}\approx0.48$ eV, and $\delta=0.319a_{C}$. The second summation in the Hamiltonian is the Hubbard interaction with $U$ being the strength of the on-site interaction, $\hat{n}_{i,s}$ is the particle number operator in lattice $i$ with spin $s$. By applying the mean field approximation, the Hubbard interaction is approximated as $U\sum_{i}(\hat{n}_{i,+}\langle\hat{n}_{i,-}\rangle+\hat{n}_{i,-}\langle\hat{n}_{i,+}\rangle)$ with $\langle\hat{n}_{i,s}\rangle$ being the particle-number expectation at lattice site $i$ with spin $s$. In the presence of hydrogen adatom, an additional Hamiltonian $H_{Adatom}$ is added to Eq. (\ref{hamiltonian}), which is defined as $H_{Adatom}=\varepsilon_{d}\sum_{s}{d^{\dag}_{s}d_{s}}+\omega_{d}\sum_{s}(d^{\dag}_{s}c_{A,s}+c^{\dag}_{A,s}d_{s})$, with $d^{\dag}_{s}$ ($d_{s}$) being the creation (annihilation) operator of the orbit at the adatom of spin $s$, $\varepsilon_{d}=7.5$ eV, and $\omega_{d}=0.16$ eV \cite{Jeongsu19}.

The Hamiltonian in mean field approximation is self-consistently  solved by iterative calculation. At the first iterative step, initial distribution of magnetic moments at the zigzag terminations are added to the model. After diagonalization of the Hamiltonian, the particle-number expectation at each lattice site is calculated as
\begin{equation}
\langle\hat{n}_{i,\sigma}\rangle=\sum_{p}f_{T}[\varepsilon(p,\sigma)]|c_{i}(p,\sigma)|^{2}
\end{equation}
where $p$ is index of energy level, $\varepsilon(p,\sigma)$ is the $p^{th}$ energy level with spin $s$, $f_{T}(\varepsilon)$ is the Fermi-Dirac distribution at temperature $T$, and $c_{i}(p,\sigma)$ is the amplitude of the wave function of energy level $\varepsilon(p,\sigma)$ at lattice site $i$. The temperature is assume to be room temperature, i.e., $T=300$ K. The particle-number expectation is insert back into the mean-field Hamiltonian for the calculation in the next iterative step. When the particle-number expectation is convergent to a stable solution, the iteration is stopped. The convergent solution is either ground state or quasi-stable excited state, which is determined by the total energy given as
\begin{equation}
E_{Total}=\sum_{p,\sigma}{f_{T}[\varepsilon(p,\sigma)]\varepsilon(p,\sigma)}-\frac{U}{2}\sum_{i}{\langle\hat{n}_{i,+}\rangle\langle\hat{n}_{i,-}\rangle}
\end{equation}
Distribution of magnetic moment is given as $M_{i}=\langle\hat{n}_{i,+}-\rangle\langle\hat{n}_{i,-}\rangle$. The total magnetic moment of the BLG QDs is $M_{Total}=\sum_{i}M_{i}$. For the BLG QDs with adatom, the total magnetic moment of the sites near to the adatom is defined as $M_{Total}=\sum_{|\mathbf{R}_{i}-\mathbf{R}_{Adatom}|<4a_{C}}M_{i}$, with $\mathbf{R}_{Adatom}$ being the coordinate of the adatom.

With different initial magnetic configuration, the iterative calculation is convergent to different solutions. For the twisted BLG QDs with varying twisting angle, the calculation procedure  follow these steps: (1) The untwisted BLG QD with a given initial magnetic configuration is calculated by the iterative calculation, and $\langle\hat{n}_{i,\sigma}\rangle$ are stored for the calculation of the next twisting step; (2) Start the next twisting step by changing the twisting angle for a small value, designated as $\Delta\theta$. (3) Update the hopping parameters $t(\mathbf{R}_{i},\mathbf{R}_{j})$ in the Hamiltonian; (4) Use the stored value of $\langle\hat{n}_{i,\sigma}\rangle$ from the convergent solution of the previous twisting step as the initial condition of the iterative calculation; (5) Iteratively calculate the solution of the updated Hamiltonian, and calculate $\langle\hat{n}_{i,\sigma}\rangle$; (6) Go back to step (2) for the next twisting step. This calculation procedure simulate the adiabatic process that the BLG QD is twisted with infinitely slow speed, because the quantum state is totally relaxed to a stable solution at each twisting step. Because the band gap of the BLG QDs has the order of magnitude at 0.1 eV, the frequency of the mechanical twisting in realistic systems (usually about 1 kHz) is much smaller than the frequency of the dynamic phase factor of the quantum state. As a result, the adiabatic approximation is valid. In our numerical calculation, $\Delta\theta=0.1^{o}$ is used. As the twisting angle change for $\Delta\theta$, the Hamiltonian only change with a small term, so that the iterative calculation in step (5) can be convergent in less than 20 steps.

\section{Numerical results and discussion}

The numerical results are analyzed in three steps. Firstly, the quantum state of the untwisted BLG QDs with varying magnetic configuration are investigated. Secondly, as the twisting angle monotonically increase in a series of twisting steps, the adiabatic evolution of the quantum state with varying initial magnetic configurations are studied. Thirdly, as the twisting angle increase to a finite value, designated as $\theta_{Max}$ after $\theta_{Max}/\Delta\theta$ twisting steps, and then decrease to zero after another $\theta_{Max}/\Delta\theta$ twisting steps, and then repeat the cycle, the bistability of the quantum state are discussed. In the numerical calculation, $N=4$ is assumed. The numerical results with different $N$ have qualitatively similar features.

\subsection{Untwisted BLG QDs}

\begin{figure*}[tbp]
\scalebox{0.73}{\includegraphics{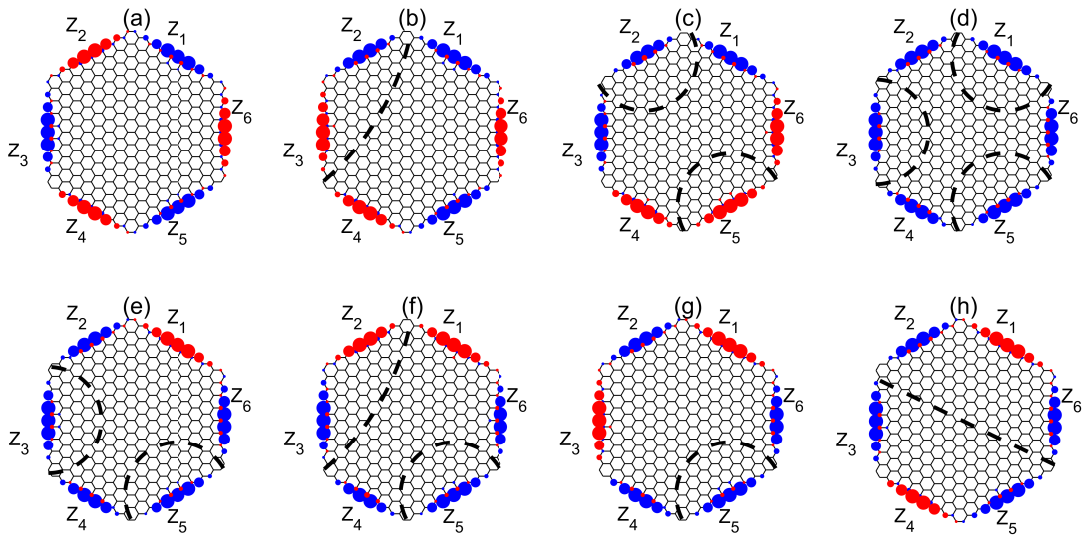}}
\caption{ Distributions of magnetic moments for the eight magnetic configurations in monolayer graphene are indicated by the size and color of the marker in each lattice site. The magnetic configurations in each sub-figure are (a) AFM, (b) AFM1, (c) AFM2, (d) FM, (e) FM1, (f) FM2, (g) FM3, and (h) FM4. The thick black dashed lines represent the domain wall of the local antiferromagnetic order. }
\label{figure_2}
\end{figure*}

We firstly review the magnetic configuration of monolayer graphene QDs with hexagonal shape and zigzag edges, and then investigate that of the BLG QDs. Because the magnetic moment at the terminations of each of the six zigzag edges can be either upward or downward, the total number of magnetic configurations are $2^{6}$. Because of time reversal symmetry, flipping the direction of magnetic moment at every lattice site generate equivalent quantum state, so that there are $2^{5}$ nonequivalent magnetic configurations. Because of the $C_{3h}$ symmetry, rotating the zigzag edge indices generate equivalent magnetic configurations. In addition, many magnetic configurations are not stable. In another words, as the iterative calculation start from these magnetic configurations, the iteration is not convergent to the solution with the same magnetic configurations. For the monolayer graphene hexagonal flake, only four magnetic configurations are stable, which are designated as ferromagnetic (FM), antiferromagnetic (AFM), AFM1 and AFM2 \cite{WeiHu19}. The spatial distributions of magnetic moments for the four magnetic configurations are plotted in Fig. \ref{figure_2}(a-d). Four more magnetic figurations ($FM1$-$FM4$) are given in Fig. \ref{figure_2}(e-h), which is not stable in monolayer graphene hexagonal flake, but exist as the magnetic configuration of one layer in the quasi-stable state of the BLG QDs.

The spontaneous magnetism has locally antiferromagnetic order, i.e. the magnetic moment at one lattice sites have opposite direction to those at the three nearest neighboring lattice sites. Thus, for bulk and semi-infinite sheet with one zigzag edge, the magnetic moment at the lattice sites belonging to sublattice A (B) are upward (downward). In zigzag nanoribbon, the zigzag terminations of the left and right edges belong to A and B sublattice, respectively. In the ground state and the quasi-stable excited state, the magnetic moments at the terminations of the two zigzag edges have opposite and the same direction, respectively. The extra energy in the quasi-stable excited state originates from the domain wall of the locally antiferromagnetic order of magnetic moment \cite{maluo2020}. At the left and right half of the zigzag nanoribbon, the magnetic moment at the lattice sites belonging to sublattice A (B) are upward (downward) and downward (upward), respective; the domain wall is along the middle axis of the nanoribbon.

In the monolayer graphene hexagonal flake with six zigzag edges, the terminating lattice sites at two adjacent zigzag edges belong to opposite sublattice. In the AFM state, the magnetic moment at the termination of two adjacent zigzag edges have opposite direction, so that the AFM state does not have domain wall, as shown in Fig. \ref{figure_2}(a). On the other hand, if the magnetic moment at the terminations of two adjacent zigzag edges have the same direction, domain wall of locally antiferromagnetic order appear along the bisect line of the corner between the two zigzag edges, which raises the total energy. In the other magnetic configurations, domain wall with varying length appear, as shown in Fig. \ref{figure_2}(b-h). As a result, the AFM state have the smallest total energy, which is the ground state. The other three quasi-stable excited states and four unstable states have larger total energy. The difference between the total energy of the quasi-stable excited states and that of the ground state is designated as exchange energy. The exchange energy is determined by the length of the domain wall. The FM3 state is obtained from the AFM state by flipping the magnetic moment at one zigzag edge (for example $Z_{5}$ in Fig. \ref{figure_2}(g)). The domain wall connects two corners at $Z_{4}-Z_{5}$ and $Z_{5}-Z_{6}$, where $Z_{i}$ represents the i$^{th}$ zigzag edge as shown in Fig. \ref{figure_2}, and $Z_{i}-Z_{j}$ represents the corner between $Z_{i}$ and $Z_{j}$. If the domain wall is strictly along the bisect lines of the two corners, which intersect at the center of the QD, the total length of the domain wall is $2R$ with $R$ being the distance from the center of the QD to one of the two corners. However, the relaxation of magnetic moments pushes the domain wall toward the zigzag edge connecting the two corners ($Z_{5}$ in this case), so that the total length of the domain wall is shorten, which in turn decreases the total energy. Meanwhile, the domain wall cannot be too close to the zigzag edge, otherwise, the magnitude of magnetic moments at the domain wall is large, which in turn increase the total energy. We can assume that the shape of the completely relaxed domain wall is a half-circle, as shown by the dashed thick line in Fig. \ref{figure_2}(g), so that the length of the domain wall is about $L_{1}=\pi R/2$. The AFM1 state is obtained from the AFM state by flipping the magnetic moment at two successive zigzag edges (for example $Z_{2}$ and $Z_{3}$ in Fig. \ref{figure_2}(b)). The domain wall connects two non-adjacent corners at $Z_{T1}-Z_{T2}$ and $Z_{T3}-Z_{T4}$. The length of the completely relaxed domain wall, designated as $L_{2}$, must be larger than $\sqrt{3}R$, which is the length of the straight line connecting the two corners, while smaller than $2R$. The FM4 state is obtained from the AFM state by flipping the magnetic moment at three successive zigzag edges (for example $Z_{6}$, $Z_{1}$ and $Z_{2}$ in Fig. \ref{figure_2}(h)). The domain wall connect the two diagonal corners, $Z_{T2}-Z_{T3}$ and $Z_{T5}-Z_{T6}$, which should be a straight line due to mirror symmetric. The length of the domain wall is $L_{3}=2R$. The domain walls in the FM3, AFM1 and FM4 states are designated as the first-, second- and third-type domain wall, which are sorted by their length. The AFM2, FM, and FM1 states have two, three, and two first-type domain walls, as shown in Fig. \ref{figure_2}(c), (d), and (e), respectively. The FM2 state has both first-type and second-type domain walls, as shown in Fig. \ref{figure_2}(f). For the four (quasi-)stable states in Fig. \ref{figure_2}(a-d), the total length of the domain wall is in ascending order, which determine the sequence of the total energy.

For untwisted BLG QDs, the magnetic configurations can be obtained by combining the magnetic configurations of the two layers. The magnetic configuration of the untwisted BLG QDs can be designated as $\Phi^{(-)}-\Psi^{(-)}-k$, where $\Phi$ and $\Psi$ represent the magnetic configurations of the bottom and top layers, respectively; $k$ represent rotation of the magnetic configuration of the top layer relative to the bottom layer for $k$ zigzag edge(s) in counter-clockwise direction; $\Phi^{-}$($\Psi^{-}$) represents reversion of direction at every zigzag edge of the corresponding magnetic configuration. $\Phi$ and $\Psi$ could be one of the magnetic configurations in Fig. \ref{figure_2}. Because the ground state of the monolayer graphene hexagonal flake is the AFM state, the ground state of the BLG QDs could be constructed by combining two layers of monolayer graphene hexagonal flake in the AFM state. In the zigzag edge of BLGs, the zigzag terminations of the top and bottom layer belong to the same sublattice, so that the low energy state favor the configuration that the magnetic moment at the two terminations have the same direction \cite{maluo21}. As a result, the ground state of the BLG QDs should be AFM-AFM-0 state. The quantum states with the other magnetic configurations are quasi-stable excited states with larger total energy. The exchange energy is increased by two factors: the presence of one pair of terminations at $Z_{Ti}$ and $Z_{Bi}$ with opposite magnetic moment increases the total energy (by the amount of which designated as $F$); the presence of domain wall with total length being designated as $L$ increases the total energy by $LW$, with $W$ being the energy of domain wall per unit length. The exchange energy is approximately equal to $nF+LW$, with $n$ being the number of pairs of terminations at $Z_{Ti}$ and $Z_{Bi}$ with opposite magnetic moment.

Numerical results confirmed that the ground state is the AFM-AFM-0 state. The spatial distribution of magnetic moments of the AFM-AFM-0 state is plotted in Fig. \ref{figure_1}(a). The magnetic moment have large magnitude in the middle of each zigzag edge. The total magnetic moment is zero. The AFM$^{-}$-AFM-0 state is obtained from the ground state by flipping the magnetic moment at the six zigzag edges of the bottom layer, so that the exchange energy is $6F$. Numerical result of the exchange energy is 0.23 eV, so that $F\approx0.038$ eV. The spatial distribution of magnetic moments of the AFM$^{-}$-AFM-0 state is plotted in Fig. \ref{figure_1}(b), which have similar feature as that of the AFM-AFM-0 state. Each of the zigzag edge at $Z_{T(2j-1)}$ and $Z_{B(2j)}$ has $2N$ terminations with downward magnetic moment, while each of the other zigzag edge has $2N+1$ terminations with upward magnetic moment, so that the total magnetic moment is upward with sizable magnitude. The first quasi-stable excited state is the AFM-FM3-3 state, which is obtained from the ground state by flipping the magnetic moment at only one zigzag edge (for example $Z_{T2}$), as shown in Fig. \ref{figure_1}(c). The exchange energy is $F+L_{1}W$, because the top layer have one first-type domain wall. Numerical result of the exchange energy is 0.11 eV, so that $W\approx0.072(eV)/L_{1}$. The exchange energy of the other quasi-stable excited state can be estimated by the formula $nF+LW$. Some of the magnetic configurations do not have stable iterative solution. For example, the AFM2-AFM2-0 state does not exist, but the iterative calculation with initial magnetic configuration being AFM2-AFM2-0 is convergent to the AFM2-AFM2-3 state. Except for the ground state and the AFM$^{-}$-AFM-0 state, the other quasi-stable excited states appear as transition quantum state in the adiabatic evolution, as discussed in the following subsections.

In the presence of adatom attaching to the top layer, the magnetic configurations can be designated as $\Phi(^{-})-\Psi(^{-})-k-M_{A}$, with the additional term $M_{A}=\pm1$ representing the direction of the magnetic moment at the adatom. The magnetic configuration of the ground state is AFM-AFM-0-$+1$, which is plotted in Fig. \ref{figure_1}(d). Near to the adatom, six lattice sites have sizable magnetic moments, which are parallel to the magnetic moment at the adatom as well as that at the other lattice sites belonging to the same sublattice. Thus, the locally antiferromagnetic order has no domain wall in the top layer. If the magnetic moment at the adatom flip, those at the lattice sites near to the adatom also flip, which in turn generate a domain wall of the locally antiferromagnetic order around the adatom. The domain wall increases the total energy. Thus, for the quasi-stable excited state with small exchange energy, the magnetic configurations are restricted in the subset $\Phi(^{-})$-AFM-$k$-$+1$. The first quasi-stable excited state is the FM3$^{-}$-AFM-0-$+1$ state, with exchange energy being 0.11 eV. Another important quasi-stable excited state for the twisting process is the AFM$^{-}$-AFM-0-$+1$ state, which is obtained by flipping the magnetic moment of the bottom layer.

\subsection{Twisted from zero to 60 degrees}

\begin{figure*}[tbp]
\scalebox{0.58}{\includegraphics{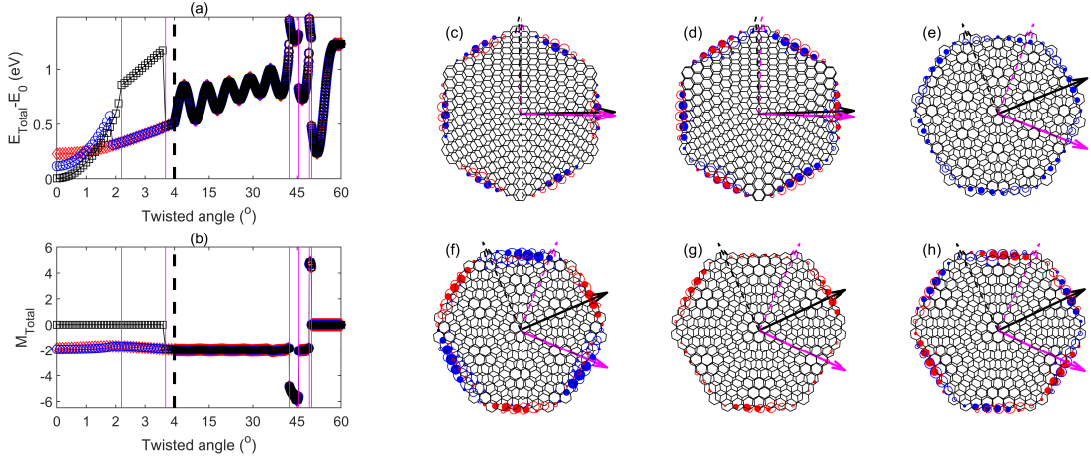}}
\caption{ For the three twisting processes of BLG QD without adatom, and with initial magnetic configuration being AFM-AFM-0, AFM-FM3-3, and AFM$^{-}$-AFM-0, the evolutions of (a) the relative total energy (total energy minus the total energy of the ground state of the untwisted BLG QD, designated as $E_{0}$), and (b) total magnetic moment, are plotted by black square, bule circle, and red diamond marked lines, respectively. The scale of the twisting angle in x-axis is stretched and contracted for the scope smaller and larger than four degree, for better visualization. The vertical thin purple lines in (a) and (b) mark the six critical twisting angles, where the distribution of magnetic moments of E-AFM-AFM-0 sharply change. (c-h) are the distribution of magnetic moments with the twisting angle barely exceeding the corresponding six critical twisting angles in (a,b). The empty and filled markers represent the magnetic moment at the lattice sites of the top and bottom layer, respectively. }
\label{figure_3}
\end{figure*}

As the twisting angle monotonic increase in a series of twisting steps, the quantum states of the whole system adiabatically evolve, so that the total energy and the spatial distribution of magnetic moments also change as functions of the twisting angle. The adiabatical evolution that starts from a particular magnetic configuration of the untwisted BLG QD (with adatom) is designated as E-$\Phi(^{-})-\Psi(^{-})-k(-M_{A})$. In this subsection, the evolutions with different initial magnetic configuration are compared. Because the twisting process usually start from the ground state, E-AFM-AFM-0 is discussed in details.

The dependent of the total energy and total magnetic moment on the twisting angle are plotted in Fig. \ref{figure_3}(a) and (b), respectively. As the twisting angle reaches a critical value at 1.8$^{o}$, the spatial distributions of magnetic moments of E-AFM-FM3-3 sharply change to be the same as that of E-AFM$^{-}$-AFM-0, so that the total energy and total magnetic moment of E-AFM-FM3-3 also sharply change to be the same as those of E-AFM$^{-}$-AFM-0. Similar phenomenon occurs for E-AFM-AFM-0. As the first critical twisting angle at 2.2$^{o}$ being reached, the spatial distributions of magnetic moments sharply change to have magnetic configuration of FM$^{-}$-FM-0, as shown in Fig. \ref{figure_3}(c), but the change of the total energy and total magnetic moment remain being smooth. The direction of magnetic moments at $Z_{T(2j-1)}$ and $Z_{B(2j)}$ is flipped, while those at $Z_{T(2j)}$ and $Z_{B(2j-1)}$ remains unchanged. At the second critical twisting angle at 3.7$^{o}$, the spatial distributions of magnetic moments sharply change to have magnetic configuration AFM$^{-}$-AFM-0, because the magnetic moments at the terminations of $Z_{T(2j)}$ and $Z_{B(2j)}$ are flipped, as shown in Fig. \ref{figure_3}(d). Thus, the following evolution sharply changes to be the same as E-AFM$^{-}$-AFM-0. More numerical results of the evolutions with difference initial magnetic configuration have similar phenomenon with different number and value of the critical twisting angles. Before the twisting angle reaches 4$^{o}$, almost all of the evolutions have sharply changed to be the same as E-AFM-AFM$^{-}$-0. As the twisting angle further increases, the spatial distribution of magnetic moments of E-AFM-AFM$^{-}$-0 undergos four more sharply change at the critical twisting angle 43.4$^{o}$, 46.5$^{o}$, 50.1$^{o}$, and 50.9$^{o}$, which are plotted in Fig. \ref{figure_3}(e), (f), (g), and (h), respectively. The quantum states with large magnitude of total magnetic moment is induced in Fig. \ref{figure_3}(e) and (g), in which the magnetic moment of all zigzag terminations are parallel.

The sharply change of the magnetic configuration is due to the competition between two factors: intra-layer domain wall and inter-layer superexchange interaction. In general, the locally antiferromagnetic order of magnetic moment has minimal total energy. In the ground state of the untwisted BLG QD, the magnetic moment at every lattice site is antiparallel to that at the intra-layer nearest neighboring sites as well as at the inter-layer nearest neighboring site (for the AB stacking sites). As the BLG QD being twisted, the AB stacking order at almost all of the lattice sites are broken. Each lattice site is neither aligned to a lattice site nor a hollow site in the other layer. Thus, the distribution of magnetic moments at the two layers does not have exactly inter-layer antiferromagnetic order (i.e. a pair of inter-layer nearest neighboring lattice sites might not have antiparallel magnetic moment), so that the superexchange energy is increased. The numerical result in Fig. \ref{figure_3}(a) confirm that the total energy is increased as the BLG QD being twisted. Flipping the magnetic moment at one zigzag edge can induce a domain wall, as shown in Fig. \ref{figure_2}, which increase the total energy. Meanwhile, the magnetic moments at the lattice sites within an area between the zigzag edge and the domain wall are flipped, which could change the inter-layer superexchange interaction. If the total number of pairs of inter-layer nearest neighboring sites with parallel magnetic moments is decreased, the superexchange interaction energy is decreased, which in turn decrease the total energy. At each twisting step, the quantum state could either evolves to a similar quantum state with the distribution of the magnetic moments slightly changing, or to another quantum state with the magnetic moments at the terminations of certain zigzag edge(s) being flipped. The evolution choose the quantum state with smaller total energy, so that the quantum state in the new twisting step is quasi-stable. As the twisting angle exceed the first critical angle of E-AFM-AFM-0 at 2.2$^{o}$, flipping the magnetic moments at the terminations of $Z_{T(2j-1)}$ and $Z_{B(2j)}$ result in a quantum state with smaller total energy, so that the sharply change of the quantum state occurs. Because of the same mechanism, at a few larger critical twisting angles, the distribution of the magnetic moments sharply changed. Because the three fold rotation symmetric of the untwisted BLG QD are preserved during the twisting, the pattern of the flipping have three fold symmetric, i.e., for each layer, magnetic moments at the terminations of the three zigzag edges with odd (or even) indices must be flipped together. If the initial distribution of magnetic moments of the untwisted BLG QD breaks the three fold rotation symmetric, such as the first quasi-stable excited state in Fig. \ref{figure_1}(c), after the twisting angle exceeding the first critical angle, the quantum state sharply changes to have a magnetic configuration with three fold rotation symmetric.

As the twisting angle reaches 60$^{o}$, the lattice structure of the BLG QDs become AA stacked. Evolutions with varying initial magnetic configurations reach the same magnetic configuration (similar to that in Fig. \ref{figure_3}(h)). Using the same marker of zigzag edges as those in Fig. \ref{figure_1}(a), the magnetic configuration is AFM-AFM-0. However, with the twisted angle being 60$^{o}$, $Z_{Ti}$ is on top of $Z_{B(i-1)}$. Thus, at each pair of zigzag edges, the magnetic moments at the top and bottom layer are antiparallel. Because of the three fold rotation symmetry, the lattice structure of the BLG QDs with twisting angle being 120$^{o}$ is the same as that of the untwisted BLG QDs. If the BLG QDs are further twisted from 60$^{o}$ to 120$^{o}$, the evolutions of the quantum states is back to the AFM-AFM$^{-}$-0 state.

\begin{figure*}[tbp]
\scalebox{0.73}{\includegraphics{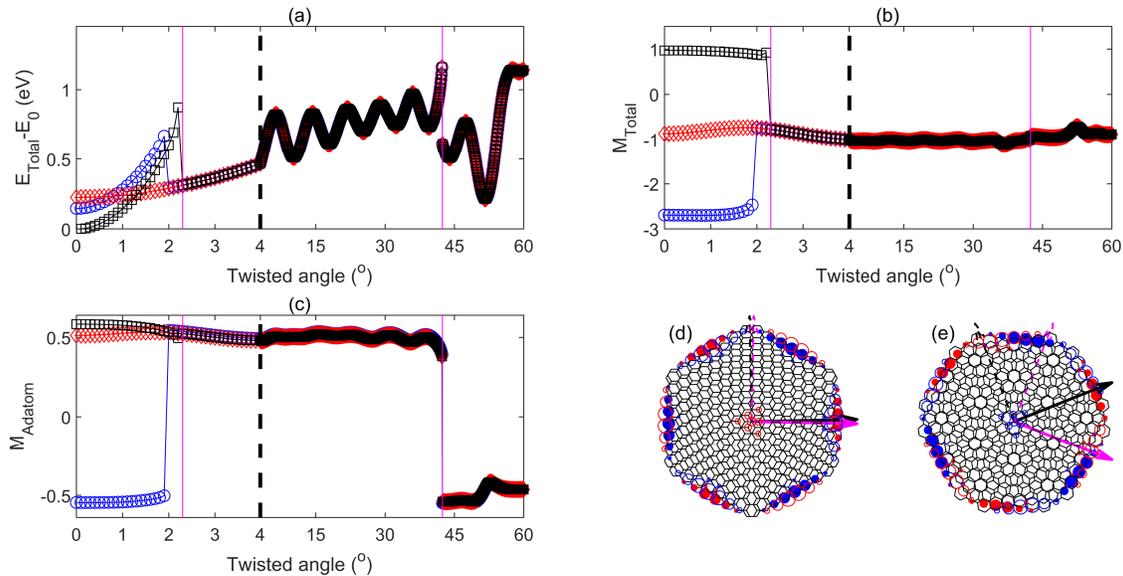}}
\caption{ For the three twisting processes of BLG QD with adatom, and with initial magnetic configuration being AFM-AFM-0-$+1$, FM3$^{-}$-AFM-0$+1$, and AFM$^{-}$-AFM-0-$+1$, the evolutions of (a) the relative total energy, (b) total magnetic moment, and (c) total magnetic moment near to the adatom, are plotted by black square, bule circle, and red diamond marked lines, respectively. The scale of the twisting angle in x-axis is stretched and contracted for the scope smaller and larger than four degree, for better visualization. The vertical thin purple lines in (a-c) mark the two critical twisting angles, where the distribution of magnetic moments of E-AFM-AFM-0-$+1$ sharply change. (d,f) are the distribution of magnetic moments with the twisting angle barely exceeding the corresponding two critical twisting angles in (a-c). The empty and filled markers represent the magnetic moment at the lattice sites of the top and bottom layer, respectively. }
\label{figure_4}
\end{figure*}

In the presence of an adatom on top of the center of the top layer, the distribution of magnetic moments also sharply change during the evolution, but the flipping of the magnetic moments at the top layer is restricted. Because the adatom locates at the center axis of the systems, the twisting does not change the position of the adatom, nor does it change the relative position of the adatom to the top layer. As discussed in the previous subsection, flipping the magnetic moment at the adatom or the zigzag edges of the top layer generates domain walls in the top layer with extra length, which in turn cost more energy. Thus, during the twisting process, the magnetic moment at the adatom and the zigzag edge of the top layer is locked, i.e., the magnetic configurations are restricted in the subsets $\Phi(^{-})$-AFM-$k$-$+1$ and $\Phi(^{-})$-AFM$^{-}$-$k$-$-1$. This restriction reduces the chance of the sharp changing of the distribution of the magnetic moments during the twisting process. The numerical results of the total energy, $M_{Total}$, and $M_{Adatom}$ during the twisting process versus the twisting angle are plotted in Fig. \ref{figure_4}(a), (b), and (c), respectively. Only two critical angles exist, after exceeding which the distributions of the magnetic moments are plotted in Fig. \ref{figure_4}(d) and (e). For E-AFM-AFM-0-+1, at the first critical angle (2$^{o}$), the evolutions sharply changed to be the same as E-AFM$^{-}$-AFM-0-+1. Similar phenomena occur for E-AFM-FM3-3-+1. At the second critical angle (42$^{o}$), the magnetic moments at both of the the adatom and the top layer flip, so that $M_{Adatom}$ flips sign.

\subsection{Hysteresis Loop}

\begin{figure}[tbp]
\scalebox{0.66}{\includegraphics{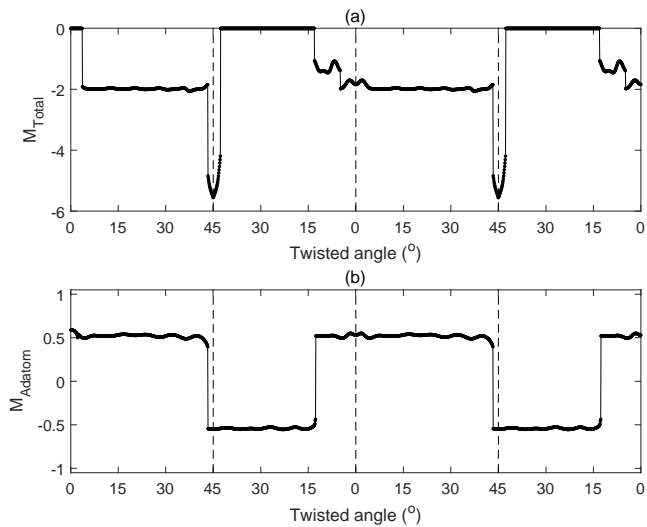}}
\caption{ The hysteretic evolutions of (a) the total magnetic moment in BLG QD without adatom, and (b) the total magnetic moment near to the adatom in BLG QD with adatom. The initial quantum state are the ground state. The twisting angle repeatedly increases from zero to 45$^{o}$ and then decrease to zero. }
\label{figure_5}
\end{figure}

In this subsection, the back and forth twisting of the BLG QDs with the twisting angle being between zero and $\theta_{Max}$ is investigated. In order to explained the hysteretic properties of the evolutions of the quantum states, an evolution with two twisting steps is described. The initial system is the twisted BLG QD with twisting angle being $\theta$. The twisting angle is changed to be $\theta+\Delta\theta$ and $\theta$ at the first and second twisting step, respectively. Designating the quantum states at the three successive twisting angles as $\Theta$, $\Theta^{\prime}$, and $\Theta^{\prime\prime}$, the iteration calculations at the first and second twisting steps start from the initial quantum state $\Theta$ and $\Theta^{\prime}$, whose convergent solutions are $\Theta^{\prime}$ and $\Theta^{\prime\prime}$, respectively. If $[\theta,\theta+\Delta\theta]$ does not cross any critical angle, $\Theta^{\prime}$ ($\Theta^{\prime\prime}$) is slightly different from $\Theta$ ($\Theta^{\prime}$), and $\Theta^{\prime\prime}$ is exactly the same as $\Theta$, i.e. the first twisting step is reversible. By contrast, if $[\theta,\theta+\Delta\theta]$ cross a critical twisting angle, $\Theta^{\prime}$ is largely different from $\Theta$. Because $\Theta^{\prime}$ is quasi-stable, as the twisting angle being tuned back to $\theta$ at the second twisting step, $\Theta^{\prime\prime}$ is slightly different from $\Theta^{\prime}$, so that $\Theta^{\prime\prime}$ is largely different from $\Theta$. Thus, the changing of the quantum state in the first twisting step is irreversible. In another words, the evolution of quantum state is dependent on the historical path of $\theta$. If $\theta_{Max}$ is smaller than the first critical twisting angle, the evolution is not hysteretic, i.e., the quantum state evolves back to the initial quantum state as $\theta$ reaches zero at each cycles. If $\theta_{Max}$ is larger than a critical twisting angle, as the twisting angle reaches zero at each cycles, the quantum state is different from the initial quantum state.

The numerical results from the previous subsection show that the first-critical twisting angle of E-AFM$^{-}$-AFM-0(-+1) is 43.5$^{o}$ (42$^{o}$) for the BLG QDs without (with) adatom. Thus, if $\theta_{Max}$ is smaller than 43.5$^{o}$ (42$^{o}$), the back and forth twisting with initial quantum state being AFM$^{-}$-AFM-0(-+1) is not hysteretic. If the initial quantum state is not AFM$^{-}$-AFM-0(-+1), before $\theta$ reaches 4$^{o}$, the quantum state sharply change to be the same as that of E-AFM$^{-}$-AFM-0(-+1). As a result, in order to obtain hysteretic evolution of the quantum state in the BLG QD without (with) adatom, $\theta_{Max}$ need to be larger than 43.5$^{o}$ (42$^{o}$). Because the quantum states with different distributions of magnetic moments have different value of total magnetic moment (and total magnetic moment near to the adatom), which can be measured in experiment, the quantum states of the BLG QDs without and with adatom are characterized by $M_{Total}$ and $M_{Adatom}$ in the following discussion.

For the BLG QD without adatom, $M_{Total}$ as a function of twisting angle in the evolution is plotted in Fig. \ref{figure_5}(a). In the first half of the first cycle, as $\theta$ increases from zero and exceeds 3.8$^{o}$, $M_{Total}$ sharply change from zero to -1.9. The following evolution is periodic. In the other words, the evolution enter the hysteresis loop at this twisting step. As the twisting angle further increases and exceeds 43.5$^{o}$, $M_{Total}$ sharply changes from -1.8 to -4.8. In the second half of each cycle, as the twisting angle decreases from 45$^{o}$ to zero, $M_{Total}$ does not sharply change at the twisting angle 43.5$^{o}$, but sharply change to zero, -1, and -2 at three other twisting angles, 42.8$^{o}$, 13.5$^{o}$ and 5.0$^{o}$, respectively. Thus, the second half of the hysteresis loop has four platforms. As the twisting angle increasing and decreasing, the functional dependence of $M_{Total}$ on $\theta$ are different from each other. At zero twisting angle (except for the initial state at the beginning of the first cycle), the quantum states evolve back to be the AFM$^{-}$-AFM-0 state.

For the BLG QD with adatom, $M_{Adatom}$ as a function of $\theta$ in the evolution is plotted in Fig. \ref{figure_5}(b). The feature of the evolution is similar to that in Fig. \ref{figure_5}(a), except that the second half of the hysteresis loop has two platforms.

\section{Conclusion}

In conclusion, the quantum states of twisted and untwisted BLG QDs with and without adatom are studied. The ground state of the untwisted BLG QDs is obtained by minimizing the superexchange interaction, which is determined by the distributions of the magnetic moments. As the twisting angle of the BLG QDs change across one of the critical angles, distribution of the magnetic moments sharply changes. For the process of increasing and decreasing twisting angle, the number and value of the critical angles are different. As a result, the back and forth twisting can drives the evolution of the distribution of the magnetic moments into a hysteresis loop. The memorial feature of the nano-structures could be use to construct nano-scale memory devices based on mechanical twisting.

\begin{acknowledgments}
This project is supported by the Natural Science Foundation of Guangdong Province of China (Grant No.
2022A1515011578), the Project of Educational Commission of Guangdong Province of China (Grant No. 2021KTSCX064), the Startup Grant at Guangdong Polytechnic Normal University (Grant No. 2021SDKYA117), and the National Natural Science Foundation of China (Grant No.
11704419).
\end{acknowledgments}

\clearpage

\end{document}